\documentclass{article} 
\usepackage{iclr2025_conference,times}

\usepackage{hyperref}
\usepackage{url}
\usepackage{booktabs}
\usepackage{graphicx}

\title{Comparison of Speech Tasks in Human Expert \\ and Machine Detection of Parkinson’s Disease}

\author{%
Peter Plantinga$^{1,2,3}$ \quad%
Roozbeh Sattari$^{1}$ \quad%
Karine Marcotte$^{4}$ \quad%
Carla Di Gironimo$^{5}$ \And%
Madeleine Sharp$^{1, 2, 6}$ \quad%
Liziane Bouvier$^{1, 2}$ \quad%
Maiya Geddes$^{1, 6}$ \quad%
Ingrid Verduyckt$^{4}$ \And%
\'Etienne de Villers-Sidani$^{1, 2, 6}$ \quad%
Mirco Ravanelli$^{3,7}$ \quad%
Denise Klein$^{1, 2, 6}$ \\ \\
1. McGill University \quad 2. CRBLM \quad 3. Mila Quebec AI Institute \quad 4. Universit\'e de Montr\'eal \\%
5. Nouvelle Voix \quad 6. Montreal Neurological Institute \quad 7. Concordia University \\
}

\iclrfinalcopy 

\begin{document}

\maketitle
\lhead{Accepted to SMASH 2025}

\begin{abstract}
The speech of people with Parkinson's Disease (PD) has been shown to hold important clues about the presence and progression of the disease. We investigate the factors based on which humans experts make judgments of the presence of disease in speech samples over five different speech tasks: phonations, sentence repetition, reading, recall, and picture description. We make comparisons by conducting listening tests to determine clinicians accuracy at recognizing signs of PD from audio alone, and we conduct experiments with a machine learning system for detection based on Whisper. Across tasks, Whisper performs on par or better than human experts when only audio is available, especially on challenging but important subgroups of the data: younger patients, mild cases, and female patients. Whisper's ability to recognize acoustic cues in difficult cases complements the multimodal and contextual strengths of human experts.
\end{abstract}

\section{Introduction}

Parkinson's Disease (PD) affects millions of people worldwide (\cite{marras2018prevalence}), but diagnosis remains challenging. Family physicians (FPs) and general neurologists (GNs) are not perfectly accurate (estimated at 76-86\% for GNs by \cite{diagnostic2014joutsa}) and extensive further testing with specialists is needed for a definitive diagnosis, causing delays and increasing the cost. One promising low-cost avenue for assisting FPs and GNs to make better initial diagnoses is speech recordings, which contain biosignals of motor symptoms as well as cognitive deficits (\cite{fang2020cognition}).

Machine learning (ML) models, especially foundation models, show promise as biomarkers for assisting clinicians with early diagnosis and monitoring due to the increased generalization capability from extensive pretraining (\cite{ali2024parkinson}). However the factors that these systems depend on to make predictions are not well-understood (\cite{Mancini2024ParkinsonSpeechExplainability}). A better understanding of the factors that both human experts and ML models rely on to understand disease-relevant signals in patient speech is sorely needed.

Our work is an early step towards understanding factors contributing to human expert and machine understanding of PD from speech. Our contributions are as follows:

\begin{enumerate}
    \item We conducted listening tests with neurologists and speech language pathologists (SLPs) with significant experience working with patients with Parkinson's disease, asking for disease prediction and reason for prediction.
    \item We trained an ML system to predict the presence of Parkinson's disease based on combining a frozen speech foundation model (Whisper) with minimal added trained parameters.
    \item We compared human experts and Whisper by looking at task-based performance, as well as performance breakdowns across demographic groups. This comparison isolates the \textit{acoustic} dimension of clinical judgment; in practice clinicians also rely on visual and patient history information that goes beyond speech alone. This comparison shows where Whisper may bring a perspective not already available to clinicians.
\end{enumerate}

We find that Whisper and human experts actually perform quite similarly across tasks and demographics, although Whisper is more accurate in a few key areas where humans are weak: younger patients, mild cases, and female patients, as well as when using spontaneous speech. This provides evidence that ML models can someday support clinicans and improve access to diagnostic care.

\section{Methods}

\subsection{Experimental Dataset (QPN)}

For our experiments with Parkinson's disease, we used a set of speech recordings from the Quebec Parkinson Network (QPN, \cite{gan2020quebec}) --- 208 patients and 52 controls. All patients and controls were recorded with a headset microphone and in a quiet room. Most patients were recorded in the ON medication state, meaning they had taken their prescribed dopaminergic treatment (e.g., levodopa) prior to the recording session. Although this can affect speech recordings, prior work has shown that dopaminergic medication has limited effects on speech production in Parkinson’s disease (\cite{cavallieri2021dopaminergic}).

The patients and controls from the QPN were asked to perform five tasks: sustained vowel phonation (SVP), repeating back sentences, reading a short passage, recalling a memory, and describing a picture (DPT). The different tasks test different motor and cognitive skills, from purely phonation and articulation (SVP) to spontaneous language production (DPT).

\subsection{Human Expert Listening Tests}

To gather feedback from human experts, we presented 64 audio samples of 30 seconds or less to a total of 7 participants who had extensive training and experience with patients with Parkinson's disease as either speech language pathologists (4) or neurologists (3). The speech samples were chosen to be carefully balanced between a number of demographic factors: patient status (PD or healthy control (HC)), subject sex (male or female), sample language (French or English), and task (listed above), with half of samples being shared between participants to estimate degree of variability in human responses.

Participants were asked to complete three items for each speech sample. Screenshots with the full text of each item are available in Appendix \ref{apx:survey}. The following is a description of the three items:

\begin{enumerate}
    \item A binary prediction about whether the sample had come from someone with
    Parkinson's Disease, or a Healthy Control
    \item A confidence rating for their prediction in the first step, out of four options: \\
    Unsure, Leaning, Confident, or Certain.
    \item A reason for their prediction in the first step, out of five options: \\
   Voice Quality, Speech Prosody, Language Use, Typical Speech, or Other (text field).
\end{enumerate}

To estimate accuracy and margin of error, we perform six trials of randomly selecting human answers where multiple are available. Because multiple responses are only available for half of all samples, we report 3 times standard deviation as our estimate of margin of error.

\subsection{Machine Learning Experiments}

We experimented with a minimal configuration of parameters on top of a frozen Whisper Small encoder in order to preserve the effects of pretraining as much as possible. We trained a small classification module consisting of the following layers: linear, attention pooling across time, linear, output (binary). The linear layers had 768 neurons and after each we added dropout at a rate of 0.2 and leaky ReLU activations.

We train our model on the set of patient and control data from QPN, excluding all patients and controls that have any samples that were reviewed by human experts. Our test set exactly matches the set of data reviewed by human experts. For training, every epoch consists of 1024 random samples of 30 seconds or less, with 32 samples per batch. The audio clips are sampled to maintain an even balance between patient status (HC or PD) and spoken task.

To reduce the incidence of overfitting on the over-sampled data, we introduced data augmentations on the waveforms. We combined two augmentations: (1) adding background noises at random signal-to-noise ratios between 0 and 15, and (2) dropping between 2 and 5 random frequencies via notch filter, and (3) dropping between 1 and 5 random chunks of the audio between 1k-2k samples each. The augmentations are dynamically applied~\cite{speechbrain}, and can only obscure important features on a fraction of the data presentations, as a regularization.

We ran six trials and recorded the average performance across trials, as well as a margin of error computed as 2 times the standard deviation.

\section{Results}

\begin{figure}
    \centering
    \includegraphics[width=\linewidth]{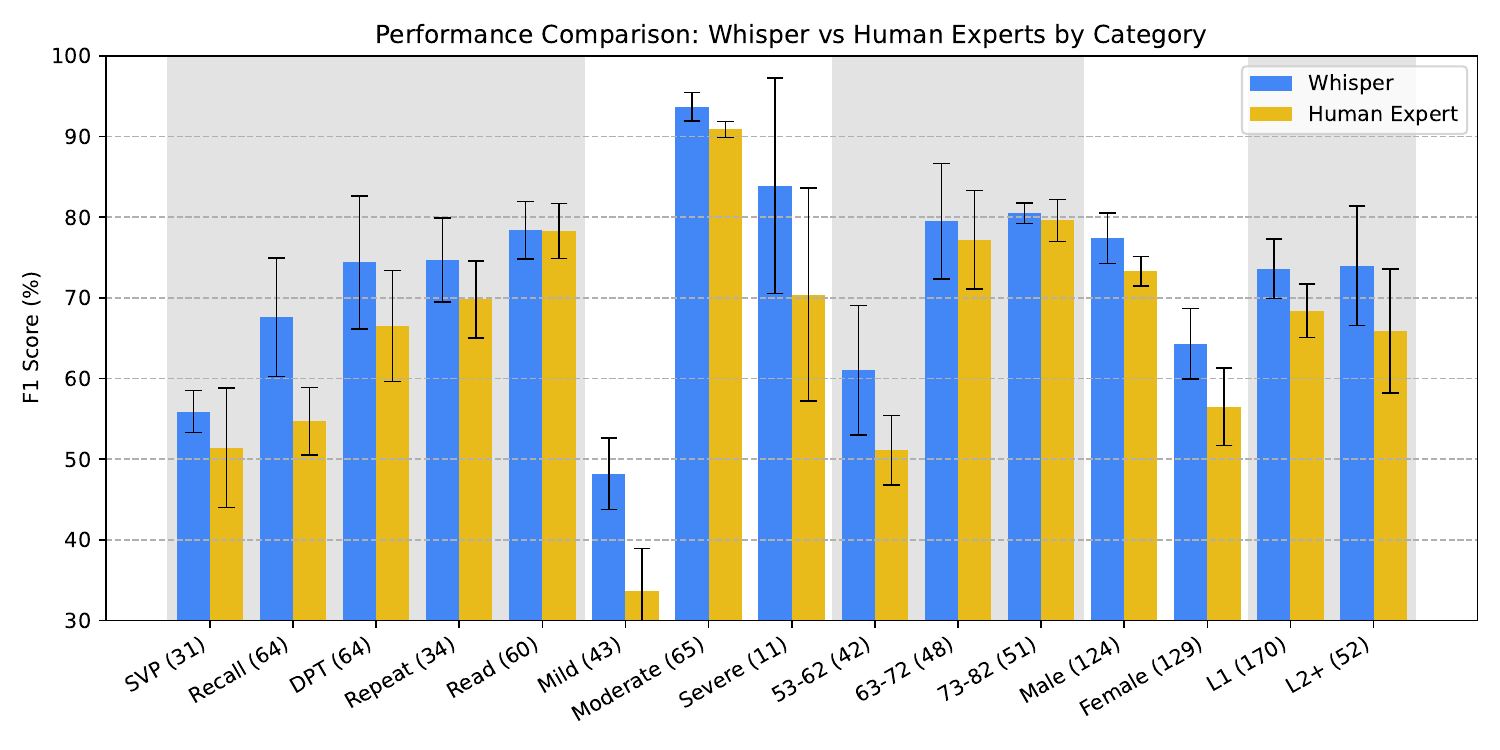}
    \vspace{-2em}
    \caption{Performance comparison of human experts and our Whisper-based system across tested categories: by task, severity, age group, sex, and language familiarity. For each category the support (number of samples) is listed in parentheses.}
    \label{fig:performance}
\end{figure}

In Figure \ref{fig:performance} we have recorded a performance comparison between human experts and Whisper across a variety of categories. The first division is by task, showing that Whisper performs better on the recall and picture description tasks. This may be related to the fact that these two tasks are spontaneous speech tasks, whereas the other tasks are more constrained.

One sign that ML models may be able to detect Parkinson's earlier than human experts comes from the results for severity and age. We find that performance is similar between human experts and Whisper, except in the youngest category (53-62 years old) and the ``mild'' category where Whisper outperforms the experts.

Finally, this figure shows comparisons by sex (male or female) and whether the sample came from a person's first language or not. The better performance on males for both human experts and Whisper seems to be robust, as we have ensured the severity and age are balanced between males and females. As for language, human experts reported more difficulty identifying Parkinson's from speech samples in a second language as they couldn't reliably determine whether irregular prosody or word-finding difficulties were from Parkinson's or the effects of speaking in a second language. This is reflected in slightly lower human expert scores for L2+, but Whisper closes the gap.

In Table \ref{tab:reason} we report the reasons that human experts gave for their decisions broken down by various categories. While language was rarely used by experts, they did use it on a few occasions for the spontaneous speech tasks -- perhaps Whisper has an advantage here, explaining better performance on these tasks. Also, Whisper shows slightly more improvement on the samples where prosody was a key factor, suggesting it may use prosody more reliably for its decisions.

\begin{table}[t]
    \centering
    \caption{Human reason for decision by task, as well as an F1 score comparison by reason.}
    \label{tab:reason}
    \vspace{1em}
    \begin{tabular}{l|ccccc|cc}
        \textbf{Reason} & \textbf{SVP} & \textbf{Recall} & \textbf{DPT} & \textbf{Repeat} & \textbf{Read} & \textbf{Human Expert} & \textbf{Whisper} \\
        \midrule
        Voice    & 95\% & 64\% & 56\% & 82\% & 58\% & $75.5\pm1.1$ & $77.0\pm3.6$ \\
        Prosody  & 5\%  & 30\% & 27\% & 18\% & 40\% & $83.4\pm4.7$ & $87.4\pm6.4$ \\
        Language & 0\%  &  7\% & 17\%  & 0\% & 2\%  & $47.0\pm37.6$ & $40.0\pm0.0$ \\
    \end{tabular}
\end{table}

\section{Discussion}

Our results suggest that machine learning models trained on speech can match or exceed the performance of clinicians when both are restricted to an artifical scenario for diagnosis based on audio recordings alone. Instead of an absolute judgement of diagnostic ability, our findings highlight complementary strengths: clinicians integrate multimodal cues and contextual knowledge that go far beyond the speech signal, while models can detect subtle acoustic irregularities that humans may overlook. On audio alone, Whisper demonstrates an advantage in difficult cases: mild, young, and female patients, as well as for spontaneous speech tasks -- which addresses the accessibility concern that not all patients may be able to read.

At the same time, accountability remains a central concern. Whereas clinicians provide explicit reasons for their judgments, the model operates as a “black box,” leaving open questions about how its predictions should be interpreted in practice. This work represents an early step toward accountability in machine learning by collecting categorical reasons for decisions that can be used for aligning model explanations. However, a great deal more work needs to be done to understand what cues models rely on, and how these cues can be used to assist clinicians.

\subsubsection*{Acknowledgments}
Supported by CRBLM (FRQNT/SC); HBHL (McGill University CFREF). Data collection provided by Roozbeh Sattari and Quebec Parkinson Network (QPN).

\bibliography{iclr2025_conference}
\bibliographystyle{iclr2025_conference}

\newpage
\appendix
\section{Appendix: Speech Rating Tool}
\label{apx:survey}

\begin{figure}[ht]
    \centering
    \includegraphics[width=\linewidth]{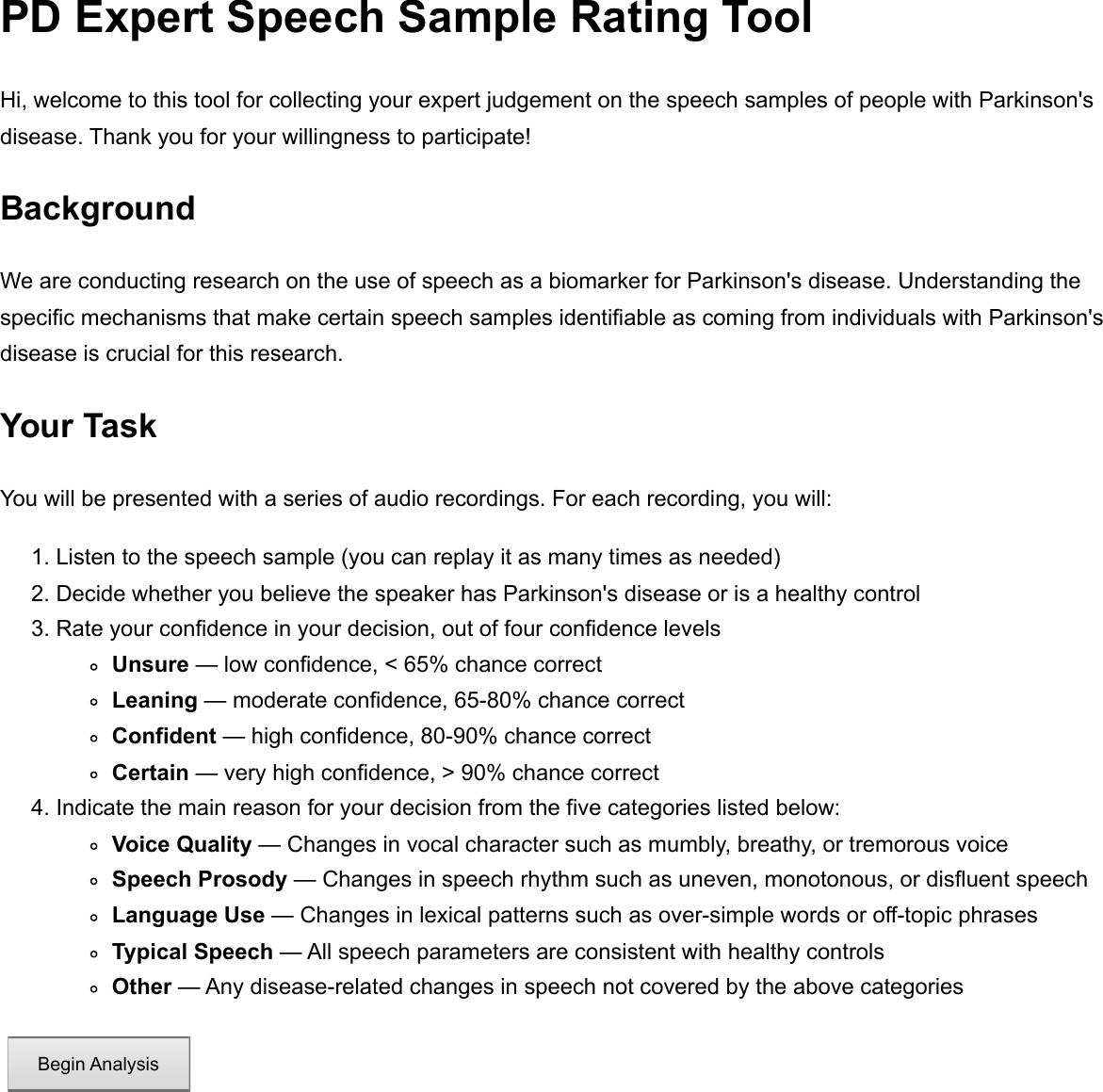}
    \caption{Screenshot of the first view in the speech sample rating tool}
\end{figure}

\begin{figure}[ht]
    \centering
    \includegraphics[width=\linewidth]{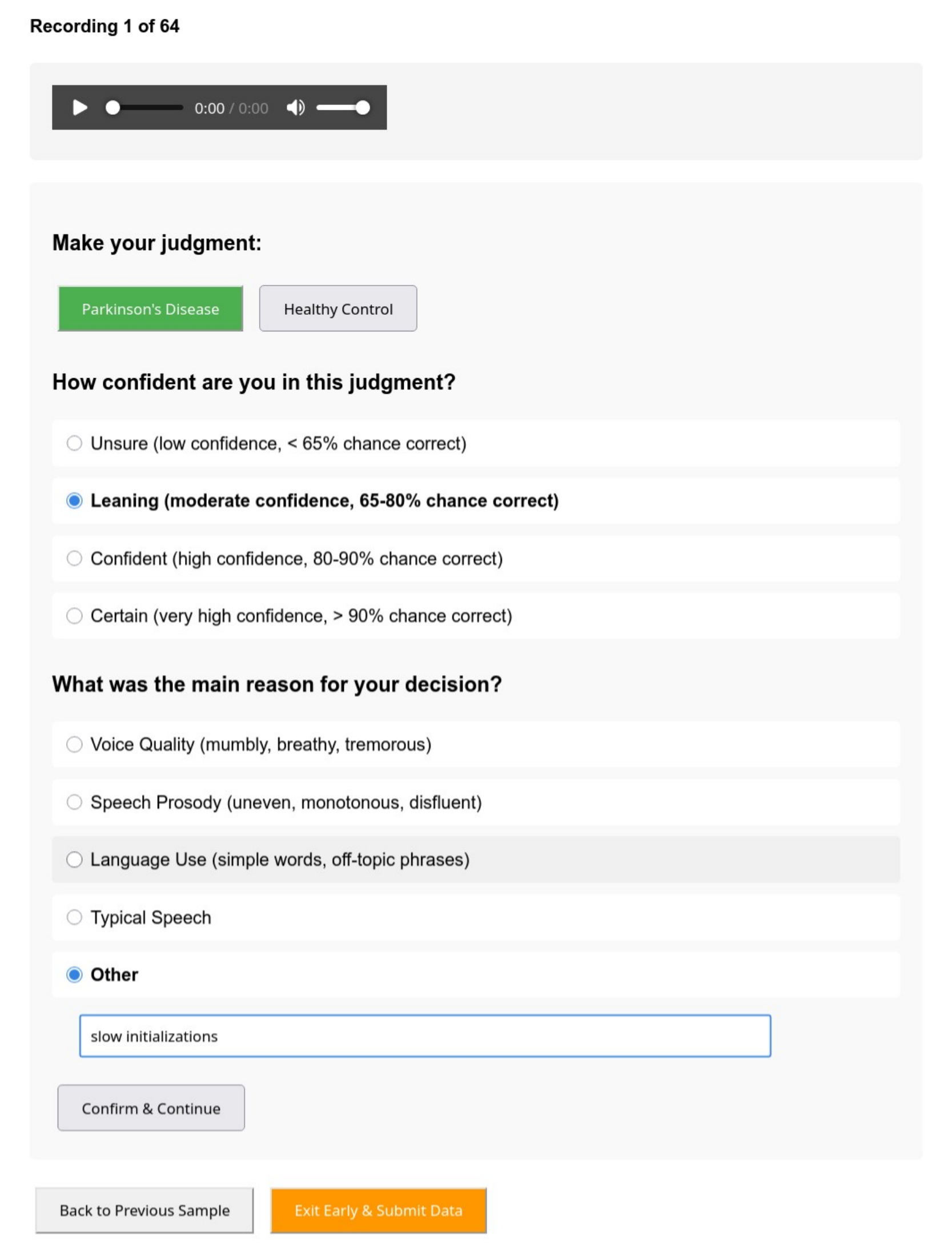}
    \caption{Screenshot of the second view in the speech sample rating tool}
\end{figure}

\begin{figure}[ht]
    \centering
    \includegraphics[width=\linewidth]{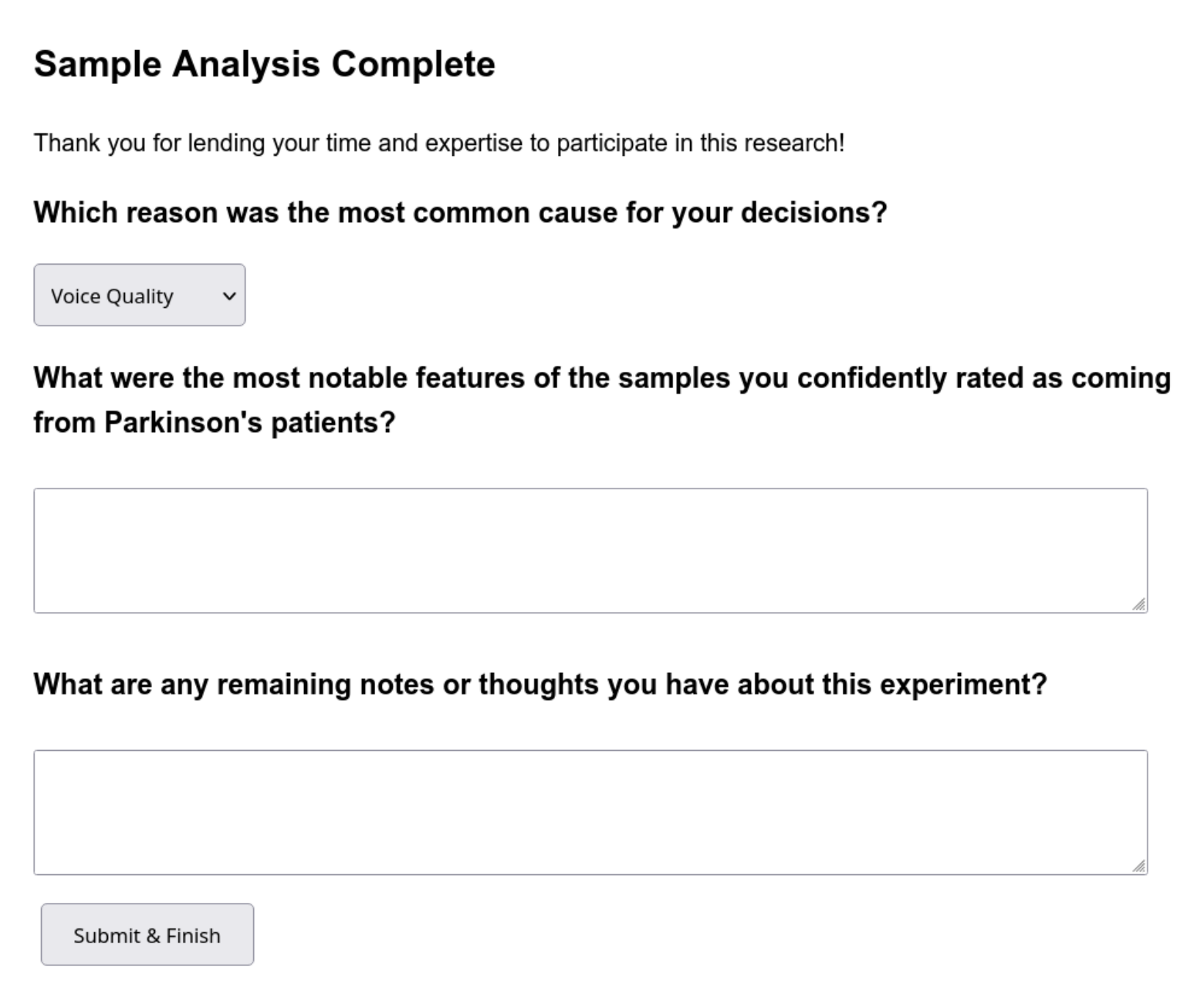}
    \caption{Screenshot of the third view in the speech sample rating tool}
\end{figure}

\end{document}